\documentclass[prb,aps,superscriptaddress,reprint]{revtex4-2}

\usepackage[T1]{fontenc}
\usepackage{graphicx}
\usepackage{units}
\usepackage{amsmath,amssymb}
\usepackage{color}
\usepackage{upgreek}
\usepackage{xcolor}
\usepackage[colorlinks=true,urlcolor=blue,linkcolor=blue,citecolor=blue]{hyperref}

\begin{document}

\title{Anisotropic supercurrent suppression and revivals in a graphene-based Josephson junction under in-plane magnetic fields}

\author{Philipp Schmidt}
\thanks{These two authors contributed equally.}
\affiliation{JARA-FIT and 2nd Institute of Physics, RWTH Aachen University, 52074 Aachen, Germany}
\affiliation{Peter Gr\"unberg Institute (PGI-9), Forschungszentrum J\"ulich GmbH, 52425 J\"ulich, Germany}
\author{Katarina Stanojević}
\thanks{These two authors contributed equally.}
\affiliation{JARA-FIT and 2nd Institute of Physics, RWTH Aachen University, 52074 Aachen, Germany}
\author{Kenji Watanabe}
\affiliation{Research Center for Electronic and Optical Materials, National Institute for Materials Science, 1-1 Namiki, Tsukuba 305-0044, Japan}
\author{Takashi Taniguchi}
\affiliation{International Center for Materials Nanoarchitectonics, National Institute for Materials Science,  1-1 Namiki, Tsukuba 305-0044, Japan}
\author{Bernd Beschoten}
\affiliation{JARA-FIT and 2nd Institute of Physics, RWTH Aachen University, 52074 Aachen, Germany}
\author{Vincent Mourik}
\affiliation{JARA Institute for Quantum Information (PGI-11), Forschungszentrum J\"ulich GmbH, 52425 J\"ulich, Germany}
\author{Christoph Stampfer}
\email{stampfer@physik.rwth-aachen.de}
\affiliation{JARA-FIT and 2nd Institute of Physics, RWTH Aachen University, 52074 Aachen, Germany}
\affiliation{Peter Gr\"unberg Institute (PGI-9), Forschungszentrum J\"ulich GmbH, 52425 J\"ulich, Germany}

\date{\today}

\begin{abstract}
We report on a tunable Josephson junction formed by a bilayer graphene ribbon encapsulated in WSe$_2$ with superconducting niobium contacts. 
We characterize the junction by measurements of the magnetic field induced interference pattern, and the AC Josephson effect manifested as "Shapiro steps", examining current dependent hysteresis and junction dynamics.
The latter can be tuned by temperature, gate voltage, and magnetic field.
Finally, we examine the evolution of the supercurrent when subjected to in-plane magnetic fields. 
Notably, we observe a strong anisotropy in the supercurrent with respect to the orientation of the in-plane magnetic field. 
When the field is parallel to the current direction, the supercurrent is suppressed, and shows revivals with increasing magnetic field, whereas it remains almost unaffected when the field is oriented in a perpendicular direction.
We suggest that this anisotropy is caused by the dependence of supercurrent interference on the junction geometry. 
\end{abstract}


\maketitle

Josephson junctions exploit the quantum mechanical phenomenon where a supercurrent flows between two superconductors separated by a thin insulating or normal conducting weak link \cite{josephson1962jul, DeGennes1964Jan, Anderson1963Mar, John1969Jan, Shepherd1972Jan}.
The unique properties of graphene provide a tunable weak link with highly transparent interfaces due to absence of Schottky barriers \cite{Giovannetti2008Jul, heersche2007mar}, while the high carrier mobility of graphene, enabling ballistic transport \cite{wang2013nov, Banszerus2016Feb}, together with its ability to host proximity induced superconductivity, make it an attractive candidate for next generation Josephson junctions~\cite{calado2015sep, Haller2022Mar}.
Graphene Josephson junctions offer several advantages over conventional junctions. Their tunability via electrostatic gating allows for dynamic control of junction properties, potentially leading to tailored and reconfigurable quantum devices~\cite{Butseraen2022Nov, schmidt2023nov, Wang2019Feb, Kroll2018Nov}.
Additionally, the implementation of a high spin-orbit material such as WSe$_2$, that gets proximity-coupled to a bilayer graphene, allows to implement Josephson junctions that potentially host topologically protected states \cite{penaranda2023apr}, which are of importance for the ongoing search for non-abelian phases of matter~\cite{Stern2010Mar,Sato2016Jun,Nayak2008Sep,Leijnse2012Nov}.

In recent years, graphene-based Josephson junctions have been intensively investigated \cite{ke2016aug,borzenets2016dec,heersche2007mar,du2008may,ojeda-aristizabal2009apr,borzenets2011sep,komatsu2012sep,mizuno2013nov,choi2013sep,li2018feb,manjarres2020feb, nanda2017jun, english2016sep, lee2018mar, park2019dec, rickhaus2012apr,draelos2018jun, zhao2020aug,gul2022jun, zhu2017feb, allen2016feb, allen2017dec, ying2020apr, rout2024jan}. 
So far, the influence of a magnetic field applied in the plane of the Josephson junction has not been systematically and angle-resolved investigated. 
However, tuning the Zeeman energy in these Josephson junctions by an in-plane magnetic field is crucial for the formation of topologically protected states \cite{kharitonov2012apr, san-jose2015dec, penaranda2023apr, xie2023oct}.

In this work, we report on a study of a tunable graphene-based Josephson junction formed by a bilayer graphene ribbon encapsulated in WSe$_2$ with superconducting niobium (Nb) contacts. 
We present a detailed characterization of the junction, which includes magnetic interference as well as Shapiro step measurements where we examine the difference between the switching currents and the damping behavior.
Furthermore, we investigate the evolution of the supercurrent when the junction is subject to in-plane magnetic fields. A strong anisotropy of the supercurrent is observed with respect to the orientation of the in-plane magnetic field, and we suggest that in-plane geometric interference effects may be its origin. 

\begin{figure*}
    \centering
    \includegraphics[width=1.0\linewidth]{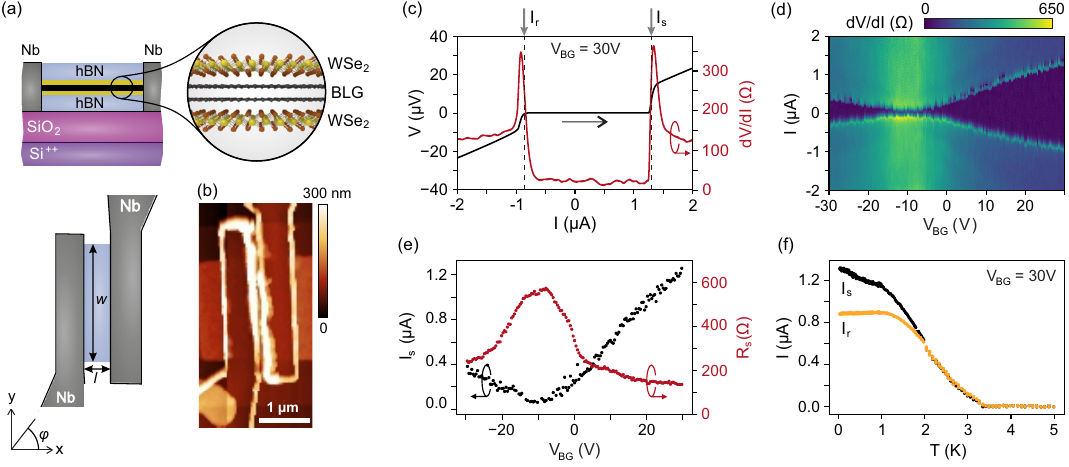}
    \caption{
    (a) Schematic of the Josephson junction. 
    A side- and top-view of the device with a length of 250\,nm and a width of 2.1\,µm is shown. 
    The BLG is located between two flakes of WSe$_2$, encapsulated in hBN and contacted by superconducting niobium (Nb) contacts.
    The van der Waals heterostructure is placed onto a Si$^{++}$/SiO$_2$ back gate.
    (b) Scanning force microscope image of the examined device showing the etched ribbon and the two niobium contacts.
    (c) DC voltage and differential resistance $dV/dI$ as a function of bias current at $V_{\mathrm{BG}} = 30$\,V. 
    The bias current was swept from negative to positive values.
    (d) Differential resistance as a function of applied back gate voltage and bias current. 
    (e) Extracted switching current $I_\mathrm{s}$ and resistance $R_\mathrm{s}$ as function of back gate voltage.
    (f)~Temperature dependent switching current $I_\mathrm{s}$ and retrapping current $I_\mathrm{r}$ at $V_{\mathrm{BG}} = 30$\,V.
    }
    \label{fig:char}
\end{figure*}

A schematic and an atomic force micrograph of our device are shown in Figs.~1(a) and (b), respectively. 
The device consists of a bilayer graphene flake symmetrically encapsulated in single layers of WSe$_2$ and thicker flakes of hexagonal boron nitride (hBN) using automated flake search \cite{uslu2024feb} and dry transfer stacking \cite{wang2013nov}.
The stack is etched into a $w=2.1\,\mathrm{\upmu m}$ wide ribbon by SF$_6$/O$_2$ reactive ion etching (RIE) through a polymethyl methacrylate (PMMA) resist mask, which has been patterned by standard electron beam lithography. 
The bilayer graphene is electrically contacted to $25\,\mathrm{nm}$ thick superconducting Nb electrodes, fabricated by RIE and consecutive sputter deposition through the same resist mask without any cleaning steps in between. 
This defines the junction length of $l=0.25\,\mathrm{\upmu m}$.
The device is placed on a highly doped silicon substrate serving as a back gate, with a 285\,nm thick separating SiO$_2$ gate dielectric.
This allows to adjust the charge carrier density of the bilayer graphene to $n=\alpha (V_{\mathrm{BG}}-V_{\mathrm{BG}}^0)$ where $\alpha_{\mathrm{BG}}$ represents the gate lever arm, which is proportional to the capacitive coupling between the back gate and the bilayer graphene. 
For more details on the fabrication procedure and a characterization of the Nb film see supplemental material sections S1 and S2~\cite{supplement}.

All measurements were conducted in a He$^3$/He$^4$ dilution refrigerator at a base temperature of 30\,mK, using a four-terminal low-frequency lock-in technique with an applied current bias, see supplemental material S3 \cite{supplement}.
In Fig.~1(c) we show a representative V-I curve as well as the differential resistance of the Josephson junction at a back gate voltage of $V_{\mathrm{BG}}=30$\,V. 
Here, the bias current is swept from negative to positive values. 
The switching current $I_\mathrm{s}$ can be identified as the switching from the superconducting to the resistive state at positive bias current, while the switching from the resistive to the superconducting state at negative bias currents defines the retrapping current $I_\mathrm{r}$.
In Fig.~1(d), the differential resistance $dV/dI$ of the device is shown as a function of the bias current $I$ and back gate voltage $V_{\mathrm{BG}}$. 
The superconducting regime (visible as the dark blue region) appears around $I=0$ for both electron doping ($V_{\mathrm{BG}}>-8$\,V) and hole doping ($V_{\mathrm{BG}}<-8$\,V), clearly distinct from the resistive regime at higher bias current values.
The extracted gate voltage dependent switching current $I_\mathrm{s}$ and switching resistance $R_\mathrm{s}$ are shown in Fig.~1(e).
The junction's $I_\mathrm{s}R_\mathrm{s}$ product ranges from $40\,\mathrm{\mu V}$ to $180\,\mathrm{\mu V}$, see supplemental material S6 \cite{supplement}.
Furthermore, we show the temperature dependence of $I_\mathrm{s}$ and $I_\mathrm{r}$ at $V_{\mathrm{BG}}=30$\,V, see Fig.~1(f). 
The temperature dependent difference between $I_\mathrm{s}$ and $I_\mathrm{r}$, which emerges below 2~K, can be understood by evaluating the junction's quality factor $Q=\sqrt{2eI_\mathrm{s}R_\mathrm{s}^2C/\hbar}$ in the framework of the resistively and capacitively shunted junction (RCSJ) model, that describes the dynamics of a Josephson junction by considering it as a parallel combination of a resistor, capacitor, and an ideal Josephson element \cite{stewart1968apr, mccumber1968jun}.
Here, the junction capacitance, estimated by fitting the RCSJ model to a measured V-I curve (see supplemental material S5 \cite{supplement} for details), is $C$=0.1\,pF and by measuring the $I_{\mathrm{s}}$ the quality factor is determined which changes from $Q>1$ to $Q<1$ at a temperature of $2.5\,\mathrm{K}$. 
Therefore, the junction dynamics changes from the underdamped to the overdamped regime where $I_\mathrm{s}$ and $I_\mathrm{r}$ become equal.
Further details on the damping and quality factor are presented in supplemental material S7 \cite{supplement}.

\begin{figure*}[t!]
    \centering
    \includegraphics[width=1.0\linewidth]{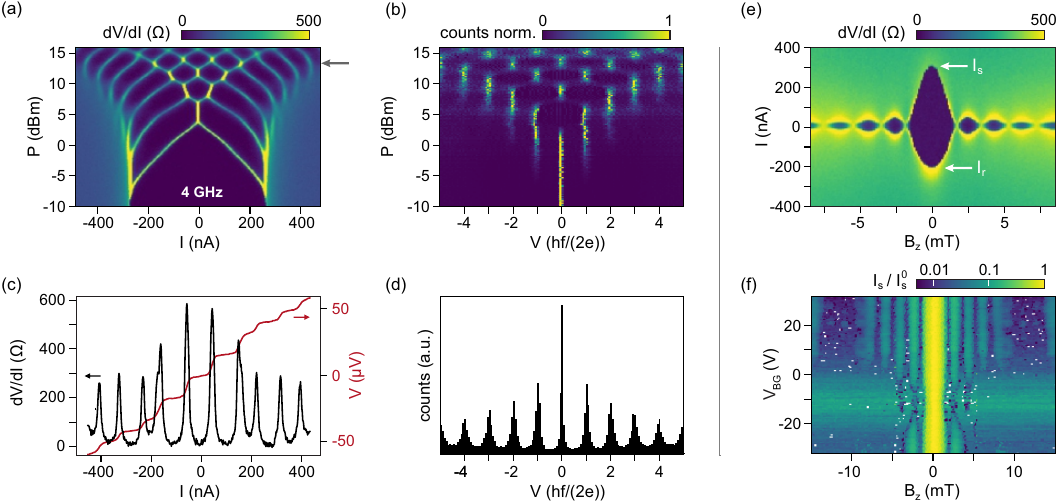}
    \caption{
    (a) Differential resistance as a function of RF signal power and bias current measured at a drive frequency of 4\,GHz and a back gate voltage of $V_\mathrm{{BG}} =30$\,V. 
    A magnetic field of $B_z=1.25$\,mT was used to decrease the switching current.
    (b) Normalized histogram of the DC voltages in units of $hf/(2e)$ as a function of RF signal power. 
    More Shapiro steps emerge with increasing power.
    (c) DC voltage (red) and differential resistance (black) as a function of bias current at an RF power of 13.8\,dBm. 
    (d)~The sum of the histogram with pronounced peaks at integer values of $V_\mathrm{{dc}}$ in units of $hf/(2e)$.
    (e)~Differential resistance as a function of perpendicular magnetic field $B_\mathrm{z}$ and bias current for $V_\mathrm{{BG}}= 0$\,V.
    (f)~Extracted normalized switching current as a function of $B_\mathrm{z}$ and back gate voltage $V_{\mathrm{BG}}$.}
    \label{fig:shapiro}
\end{figure*}

\begin{figure*}[t!]
    \centering
    \includegraphics[width=0.99\linewidth]{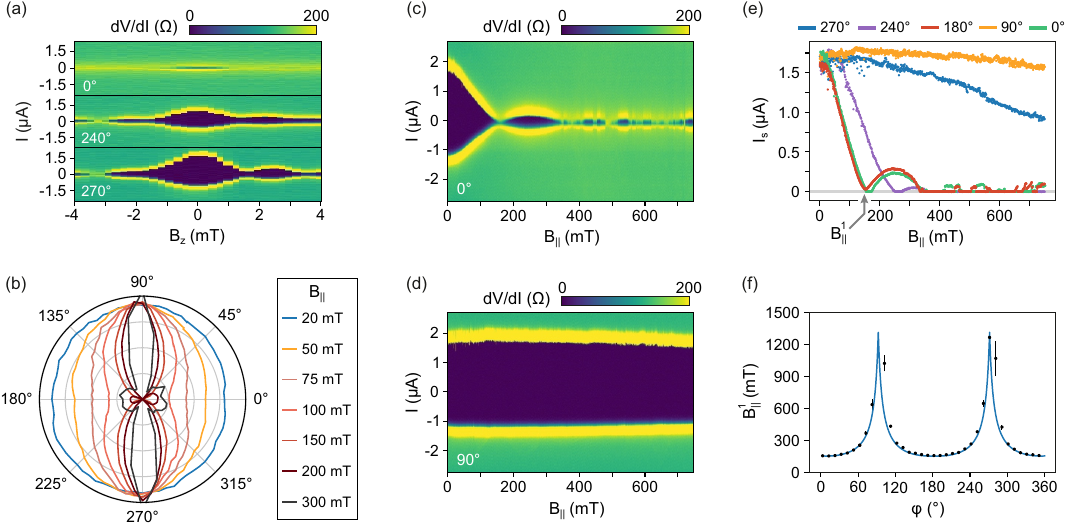}
    \caption{
    (a) Differential resistance as function of current and out-of-plane magnetic field for $B_{||}=150\,\mathrm{mT}$ at in-plane angles $\varphi=0^{\circ}$, $240^{\circ}$ and $270^{\circ}$.
    (b) Polar representation of the maximum of $I_\mathrm{s}$, extracted from the interference measurements, as a function of $\varphi$ for varying $B_{||}$.
    (c)-(d) Differential resistance as function of current and $B_{||}$ applied in the direction of the current ($\varphi=0^{\circ}$) and perpendicular $\varphi=90^{\circ}$.
    (e) Extracted switching current as a function of $B_{||}$ for various in-plane angles $\varphi$.
    The gray bar indicates the region of non-detectable switching current.
    (f)~In-plane magnetic field of the first supercurrent minimum $B^1_{||}$ for different angles $\varphi$. 
    Solid gray line represents a fit to the geometrical model described in the text. 
    An outlying point at $90^{\circ}$ at around 3\,T is not included. 
    See supplementary material section S10 \cite{supplement} for the extraction (for minima up to 300 mT) or extrapolation procedure (for minima beyond 300 mT). 
    }
    \label{fig:Bip}
\end{figure*}

To study the junction dynamics in more detail, we investigate the influence of microwave radiation on the V-I curve of the junction. 
Applying microwave radiation leads to the AC Josephson effect, which is manifested in additional plateaus in the DC voltage \cite{shapiro1963jul}, known as Shapiro steps. 
These steps occur at integer multiples of $hf/2e$ and result from the phase locking across the junction to the external frequency. 
Here, $h$ is the Planck constant, $f$ the frequency of the radio frequency (RF) signal source, and $e$ the elementary charge.
The formation of Shapiro plateaus with increasing microwave power is visible in Fig.~\ref{fig:shapiro}(a) where the measurement was taken at $V_{\mathrm{BG}}=30\,V$ and $f=4\,\mathrm{GHz}$. 
Here, the differential resistance is shown as a function of applied bias current $I$ and the applied signal power. 
The pattern qualitatively resembles the expectation for weakly damped Josephson junctions where the extension of the steps surpass $I_\mathrm{s}$ and superconducting pockets develop into the normal conducting region.
We also observe broad regions of microwave power where the resistive transition is an extended line instead of single points expected for the Bessel function behavior of overdamped Josephson junctions \cite{tinkham2004}.
This behavior has also been observed in other graphene Josephson junctions \cite{larson2020oct, kalantre2020apr, Vignaud2023Dec} and may be explained by the junction being underdamped and having a high plasma frequency.
The first Shapiro plateau starts to develop at -7\,dBm, whereas at higher powers many different plateaus emerge separated by sharp peaks in the d$V$/d$I$ curve as depicted in Fig.~\ref{fig:shapiro}(c). 
Moreover, the quantization of these steps can be seen in the power dependent histogram of the DC voltages in Fig.~\ref{fig:shapiro}(b) where the steps emerge at integer multiples of $hf/2e$. 
This is also reflected in sharp peaks of the sum histogram, see Fig.~\ref{fig:shapiro}(d).
All integer Shapiro steps are present without any sub-integer steps appearing, unlike what is observed in other two-dimensional Josephson junctions \cite{huang2023jun, lee2015jan}, suggesting that the current phase relationship is not strongly skewed. 
Additional Shapiro step measurements are shown in supplemental material S8 \cite{supplement}.

Next, we examine the magnetic field dependence of the switching current for both out-of-plane and in-plane directions. 
We start with out-of-plane magnetic fields and measure the differential resistance vs both the current $I$ and the out-of-plane magnetic field $B_\mathrm{z}$. 
The phase difference between the two superconductors induced by the magnetic field leads to a modulation of the switching current $I_\mathrm{s}$ \cite{rowell1963sep}, as illustrated in Fig.~\ref{fig:shapiro}(e).
Analyzing the periodicity of the modulation pattern leads to an effective junction length of 603\,nm and a magnetic penetration depth of $\lambda \approx175\,\mathrm{nm}$, with the area of the weak link determined from atomic force microscopy measurements, see supplemental material S4 \cite{supplement} for details.
The resulting penetration depth is in reasonable agreement with $\lambda_{\mathrm{Nb}}=150\,\mathrm{nm}$ reported for niobium \cite{gubin2005aug}.
We note that the oscillation period of the supercurrent modulation pattern remains unchanged irrespective of the applied gate voltage, see Fig.~\ref{fig:shapiro}(f).
This is in contrast to previous work on BLG/WSe$_2$ where a $2\Phi_0$ signature has been observed in the interference pattern at charge neutrality~\cite{rout2024jan}.

In Fig.~\ref{fig:shapiro}(e), a pronounced difference between the switching current and the retrapping current can be seen around the central lobe. 
This behavior can again be explained by a tunable quality factor $Q$ as the magnetic field induced modulation of the switching current leads to a transition of the quality factor from $Q \approx 3$ at the central lobe to $Q \approx 1$ at higher lobes, accompanied by a transition from the underdamped to the hysteresis-free junction dynamics.

We now analyze the junction's behavior for an in-plane magnetic field ($B_{||}$) varying the angle ($\varphi$), which is defined in Fig.~\ref{fig:char}(a).
First, the $B_{\mathrm{z}}$-induced interference pattern is shown in Fig.~\ref{fig:Bip}(a) at $B_{||}$=150\,mT applied at different angles. 
When the in-plane magnetic field is oriented parallel to the direction of the current ($\varphi=0$° and $\varphi=180$°), the supercurrent is most strongly suppressed (see upper panel in Fig.~\ref{fig:Bip}(a)). 
However, it reappears at skewed angles, such as 240°, and reaches a maximum when the field is perpendicular to the current ($\varphi=270$°).
This behavior is summarized in a polar plot of the maximum switching current  plotted as a function of the in-plane angles and the amplitude of $B_{||}$ in Fig.~\ref{fig:Bip}(b). 
With increasing amplitude of $B_{||}$, the supercurrent shows an increased anisotropy. 
While $I_{s}$ is largely unaffected when $B_{||}$ is oriented perpendicular to the current direction at $\varphi=90$° and 270°, it is suppressed when the field is parallel to it. 
It becomes zero near $B_{||}$=150\,mT and even reappears at larger field amplitudes. Additional data on the evolution of the interference pattern subject to in-plane magnetic fields is presented in the supplemental material section S9 \cite{supplement}.

Sweeping $B_{||}$ for different in-plane angles, as shown in Fig.~\ref{fig:Bip}(c), reveals not only a significant anisotropy in the switching current but also supercurrent revival effects depending on the orientation of the magnetic field. 
This is in stark contrast to the configuration where the magnetic field is oriented perpendicular to the current direction for $\varphi=270$° (see Fig.~\ref{fig:Bip}(d)), where the supercurrent only decreases by approximately 200\,nA over the entire range of the magnetic field.
This anisotropy is further illustrated in Fig.~\ref{fig:Bip}(e) by showing the extracted maximum supercurrent for various angles.
For angles perpendicular to the supercurrent flow direction, a monotonous decrease of $\approx 15 \%$  throughout the investigated field range is observed, while tilting the in-plane magnetic field towards the supercurrent flow direction leads to a progressively stronger suppression of the supercurrent, with the field of the first minimum eventually reaching a smallest value of 150 mT.

To further analyze this anisotropy, we extract the in-plane magnetic field value $B_{||}^1$, corresponding to the first minimum of $I_\mathrm{s}$ (see supplemental material S10 \cite{supplement} for a discussion and details) and plot it as a function of $\varphi$, see Fig.~\ref{fig:Bip}(f).
Assuming a finite effective thickness of the junction $d_{\mathrm{eff}}$, the in-plane magnetic field induces a magnetic flux $\Phi_{||}=B_{||} d_{\mathrm{eff}} (|w\cos{(\varphi)}| + |l\sin{(\varphi)}|)$. 
In case supercurrent suppression and its anisotropy are caused by interference effects, the first supercurrent minimum appears at magnetic fields of order $B_{||}^1=\Phi_0 / (d_{\mathrm{eff}} (|w\cos{(\varphi)}| + |l\sin{(\varphi)}|))$.
A fit of this model to the extracted data results in an effective thickness of $d_{\mathrm{eff}}=6.3\,\mathrm{nm}$.

Supercurrent interference under in-plane magnetic fields can be understood in a qualitative microscopic picture. 
In a perfectly homogeneous two-dimensional Josephson junction, supercurrent is carried by identical Andreev bound states (ABS) in many identical parallel transport channels.
Here, magnetic flux is threaded through the out-of-plane orbital component of the ABS wave function for any in-plane angle $\varphi$ (in principle allowing for interference for any value of $\varphi$), but due to the identical orbital wave functions, no interference occurs. 
However, such an idealized picture is not realistic, as it ignores microscopic disorder that will cause the many ABS present to have, at best, similar, but not identical, orbital wave functions. This will lead to the build-up of phase differences between the individual ABS under in-plane magnetic field, and the total supercurrent integrated over the entire junction will display an averaged orbital interference effect.

Possible disorder mechanisms in our devices are residual geometric disorder after encapsulating bilayer graphene between WSe$_2$ and hBN, and electrostatic disorder, for example due to defects and contaminants at the various interfaces in the layer stack \cite{dvir2021mar, Fyhn2020Jul}. 
Additionally, a narrow section of the junction in the contact area may be more strongly disordered, which may enhance supercurrent interference effects.

Beyond disorder enabling supercurrent interference effects, the assumption of a homogeneous in-plane magnetic field is not realistic. 
Flux focusing effects caused by the presence of superconducting contacts may allow for small out-of-plane components of the magnetic field, which are expected to be highly dependent on the exact device geometry. 
In an earlier work \cite{Suominen2017Jan} the anisotropy of in-plane supercurrent interference was fully attributed to flux focusing. 
Further texture may be added to the magnetic field by irregularities in the shape of the superconducting contacts, and the possibility of vortices also entering the niobium film under in-plane magnetic fields.

Our experiment cannot discriminate between the possible causes of supercurrent interference. 
Assuming such effects are significant, we expect supercurrent interference for any value of $\varphi$, and a geometric dependence on $\varphi$, rationalizing our phenomenological fitting procedure. 
We conjecture that $d_{\mathrm{eff}}$ introduced above is a phenomenological parameter that absorbs microscopic disorder effects, flux focusing, and other magnetic field inhomogeneities into a single fitting parameter. 
Its value of 6.3\,nm is about an order of magnitude larger than the thickness of the bilayer graphene, which means that the effective transverse area of the junction through which flux can be threaded is enlarged by the various possible causes of supercurrent interference.

Relating $d_{\mathrm{eff}}$ to a microscopic model and assessing its value, while disentangling the effect of flux focusing, requires modeling efforts beyond the scope of this work.
Such modeling could rely on introducing a randomization of the junction geometry in an otherwise standard classical approach following \cite{tinkham2004}. 
Alternatively, a quantum mechanical electronic transport approach could be followed, extending such modeling carried out for quasi-one-dimensional nanowire-based Josephson junctions \cite{zuo2017nov} to two-dimensional Josephson junctions. 
Any such modeling should be approached with caution due to risks of overfitting the data by introducing additional modeling parameters when attempting to create more realistic models. 

Alternative mechanisms predicting decay and revival of supercurrent under in-plane magnetic fields rely on spin physics stemming from the semiconducting band structure incorporated into the effective Hamiltonian of the Josephson junction. 
Such proposed effects include $0-\pi$ transitions of the junction's ground state due to the Zeeman effect \cite{hart2017jan, li2019jul}, similar transitions but to arbitrary phase differences $\phi_0$ (so-called $\phi_0$-junctions) \cite{sickinger2012sep, szombati2016jun} due to additional spin-orbit interaction, or more exotic physics such as topological phase transitions. 
However, for significant spin splitting to occur, which is a prerequisite for such effects, much larger magnetic fields of the order of several Teslas are anticipated in bilayer graphene. 
We observe supercurrent minima already at values as low as 150 mT, rendering spin-physics related explanations implausible. 

In summary, we have presented a tunable lateral Josephson junction consisting of bilayer graphene encapsulated in WSe$_2$. 
We have been able to tune the junction quality factor and thus its damping regime by external parameters, such as back gate voltage, magnetic field and temperature. 
This is evident from the magnetic field induced modulation of the switching current and the retrapping current. 
Furthermore, we see well-defined Shapiro steps under RF driving of the junction.
We observe a highly anisotropic suppression and revival of the supercurrent when the Josephson junction is subject to in-plane magnetic fields. 
We suspect that this anisotropic behavior is caused by orbital interference of the supercurrent.
Further research on quasi two-dimensional Josephson junctions at finite in-plane magnetic fields is required to obtain an in-depth understanding in these systems. 
We caution against narratives that fail to consider interference effects when invoking supercurrent suppression and revival to support claims of observing novel spin phenomena originating from the semiconducting band structure.\\

\begin{acknowledgments}
We thank L. Banszerus, S. Anupam and S. M Frolov for insightful discussions. 
This project has received funding from the Deutsche Forschungsgemeinschaft (DFG, German Research Foundation) under Germany’s Excellence Strategy – Cluster of Excellence Matter and Light for Quantum Computing (ML4Q) EXC 2004/1 – 390534769, from the European  Research Council (ERC) under the European Union’s Horizon 2020 research and innovation programme (grant agreement No.820254), and the Helmholtz Nano Facility \cite{albrecht2017may}. 
K.W. and T.T. acknowledge support from the JSPS KAKENHI (Grant Numbers 21H05233 and 23H02052) and World Premier International Research Center Initiative (WPI), MEXT, Japan.\\
\end{acknowledgments}

\textbf{Data availability} The data supporting the findings, as well as simulation and data processing code are available in a Zenodo repository under accession code~\cite{zenodo}.\\

\textbf{Author contributions}
P.S., V.M. and C.S. conceived this experiment. 
P.S. and K.S. fabricated the device, performed the measurements and analyzed the data. 
P.S. performed the simulation. 
K.W. and T.T. synthesized the hBN crystals. 
B.B., V.M. and C.S. supervised the project. 
P.S., K.S. and V.M. wrote the manuscript with contributions from all authors. 
P.S. and K.S. contributed equally.

\end{document}


\title{\bf Supplemental Material for: \\ Anisotropic supercurrent suppression and revivals in a graphene-based Josephson junction under in-plane magnetic fields}

\author{Philipp Schmidt}
\thanks{These two authors contributed equally.}
\affiliation{JARA-FIT and 2nd Institute of Physics, RWTH Aachen University, 52074 Aachen, Germany}
\affiliation{Peter Gr\"unberg Institute (PGI-9), Forschungszentrum J\"ulich GmbH, 52425 J\"ulich, Germany}
\author{Katarina Stanojević}
\thanks{These two authors contributed equally.}
\affiliation{JARA-FIT and 2nd Institute of Physics, RWTH Aachen University, 52074 Aachen, Germany}
\author{Kenji Watanabe}
\affiliation{Research Center for Electronic and Optical Materials, National Institute for Materials Science, 1-1 Namiki, Tsukuba 305-0044, Japan}
\author{Takashi Taniguchi}
\affiliation{ Research Center for Materials Nanoarchitectonics, National Institute for Materials Science,  1-1 Namiki, Tsukuba 305-0044, Japan}
\author{Bernd Beschoten}
\affiliation{JARA-FIT and 2nd Institute of Physics, RWTH Aachen University, 52074 Aachen, Germany}
\author{Vincent Mourik}
\affiliation{JARA Institute for Quantum Information (PGI-11), Forschungszentrum J\"ulich GmbH, 52425 J\"ulich, Germany}
\author{Christoph Stampfer}
\affiliation{JARA-FIT and 2nd Institute of Physics, RWTH Aachen University, 52074 Aachen, Germany}
\affiliation{Peter Gr\"unberg Institute (PGI-9), Forschungszentrum J\"ulich GmbH, 52425 J\"ulich, Germany}

\maketitle

\section{Sample fabrication}
Thin flakes of hexagonal boron nitride (hBN), bilayer graphene (BLG), and tungsten diselenide (WSe$_2$) were mechanically exfoliated. 
The identification of these thin flakes was achieved using optical contrast under a microscope, verified by a machine learning approach that considers thickness-dependent optical contrast and the geometric shape of the flake \cite{uslu2024feb}. 
Using these flakes, van der Waals heterostructures are assembled by a polymer-based dry transfer technique \cite{wang2013nov} and deposited on a Si$^{++}$ substrate covered with 285\,nm of SiO$_2$ which serves as a back gate. 
The stack is etched into a rectangle by SF$_6$ reactive ion etching (RIE) through a poly(methyl methacrylate) resist mask, which has been patterned by standard electron-beam lithography. 
Then, bond pads and normal conducting leads (connecting the bond pads and the superconductors) consisting of 70\,nm gold with a 5\,nm chromium adhesion layer were patterned and deposited by electron beam evaporation. 
The BLG is electrically contacted by SF$_6$ RIE etching and subsequent sputter deposition of 25\,nm niobium with a 5\,nm titanium adhesion layer through the same resist mask.
Finally, a $\approx 5 \,\mathrm{nm}$ thin hBN flake is placed on the device and a Cr/Au top gate is deposited on the entire device. This gate was not able to tune the junction's charge carrier density, which we suspect is caused by a relatively large distance to the junction due to tall fencing effects from sputtering Niobium onto a lithographic lift-off mask. We kept the gate at $0\,\mathrm{V}$ for the remainder of the experiments. We note that, although this top gate was not used in the study, it leads to a high device capacitance in a RCSJ-model context due to the small distance of the top gate to the superconducting contacts.

\section{Characterization of the niobium film}
\label{sec:Nb}

\begin{figure}[h]
    \centering
    \includegraphics[width=0.99\linewidth]{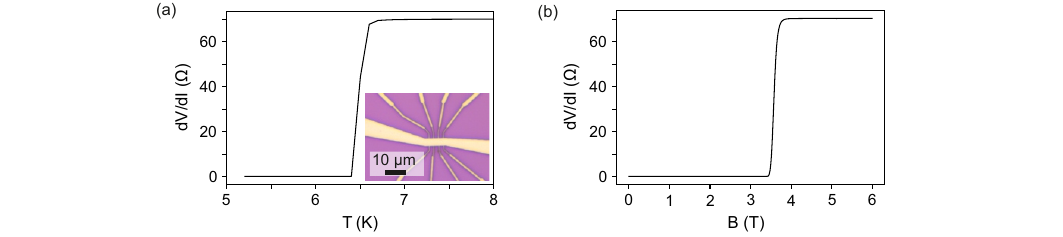}
    \caption{(a) Differential resistance as function of temperature of a 25\,nm niobium film with a 5\,nm thick titanium adhesion layer deposited on Si/SiO$_2$. 
    The critical temperature is approximately 6.4\,K. 
    Inset: Optical micrograph of the reference structure. 
    (b) Differential resistance as a function of magnetic field. 
    The critical field is 3.5\,T.}
    \label{fig:Nb}
\end{figure}

To characterize the superconducting material used for the Josephson junction, the critical temperature and critical magnetic field of a nominally identical niobium reference structure is measured, see Fig.~\ref{fig:Nb}.
The temperature-dependent resistance (Fig.~\ref{fig:Nb}(a)) shows a critical temperature of $T_{\mathrm{c}}=6.5\,\mathrm{K}$. 
The out-of-plane critical field is determined to be 3.5\,T (see Fig.~\ref{fig:Nb}(b)).

\clearpage

\section{Experimental setup}
\label{sec:setup}

\begin{figure}[h]
    \centering
    \includegraphics[width=0.99\linewidth]{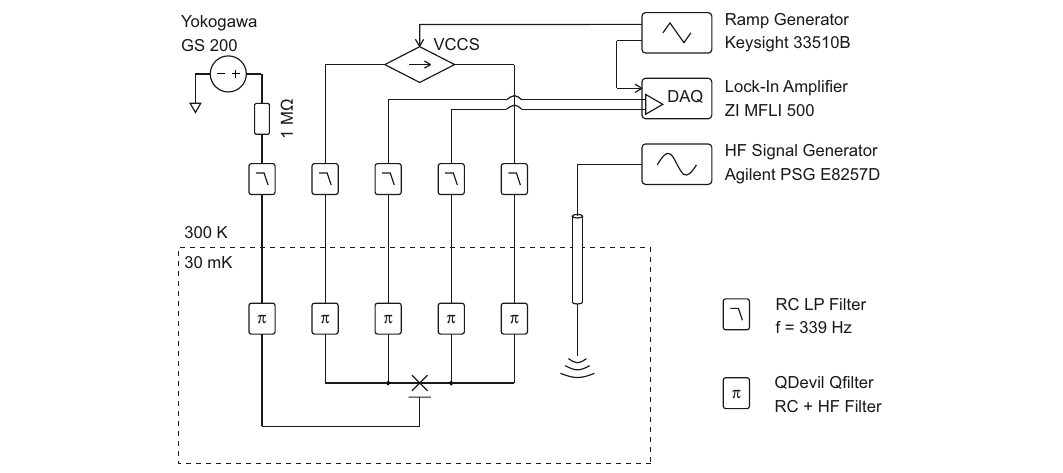}
    \caption{Schematic of the measurement setup. 
    The Josephson junction device is located inside a dilution refrigerator at a base temperature of 30\,mK. 
    All lines leading into the fridge are equipped with QDevil Qfilters located at the mixing chamber plate with a base temperutre of  30\,mK. 
    Further, RC low-pass (LP) filters with a cutoff frequency of 339\,Hz are connected to all lines outside the fridge. 
    The back gate voltage is applied using a Yokogawa GS 200 inline with an 1\,M$\mathrm{\Omega}$ resistor. 
    A triangle bias voltage generated by a ramp generator Keysight 33510B is used as a source for the self built voltage-controlled current source (VCCS) \cite{volmer2022jul}. 
    This bias current is applied to the junction. 
    Using a lock-in amplifier ZI MFLI 500, the four-point voltage is measured as well as the voltage applied to the VCCS. 
    To measure the AC Josephson effect, an Agilent PSG E8257D is used as a radio frequency (RF) source which is applied to a ground plane antenna close to the sample. }
    \label{fig:setup}
\end{figure}

The device was measured in a dilution refrigerator at a base temperature of 30\,mK. 
The schematic wiring is shown in Fig.~\ref{fig:setup}.
To reduce the electron temperature and improve the signal-to-noise ratio, all wiring inside the fridge is equipped with cryogenic low-pass filters (QDevil Qfilter) at the mixing chamber stage and RC low-pass (LP) filters with a cutoff frequency of 339\,Hz at room temperature.
Using a Keysight 33510B, triangular-shaped voltage pulses are applied to a home-built voltage-controlled current source (VCCS) \cite{volmer2022jul} to drive a constant current through the sample.
A Zurich Instruments (ZI) MFLI lock-in amplifier is used to apply a small AC excitation and measure the differential voltage and  DC-voltage of the junction, as well as the triangular bias voltage controlling the VCCS.
The back gate voltage is applied using a Yokogawa GS 200 inline with a 1\,M$\mathrm{\Omega}$ resistor.
Additionally, an Agilent PSG E8257D is connected to a $\lambda/4$-antenna near the sample. 
By applying a radiofrequency (RF) signal, the AC Josephson effect can be investigated.

\begin{figure}[h]
    \centering
    \includegraphics[width=0.99\linewidth]{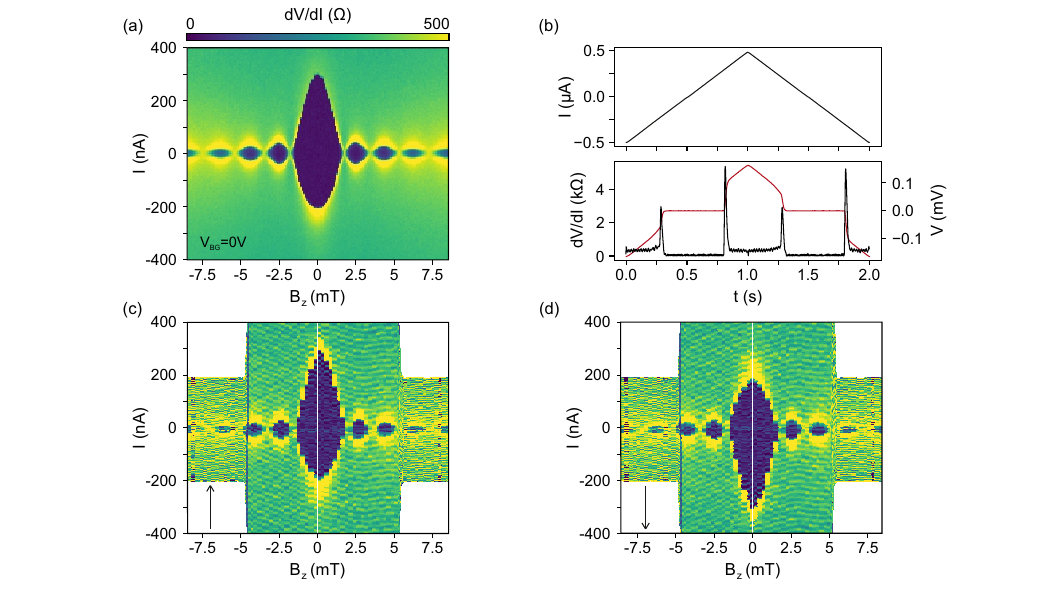}
    \caption{(a) Measured interference pattern for $V_\mathrm{{BG}}$=0\,V using a point by point - wise data acquisition technique.
    (b) Visualization of the fast measurement technique for a back gate voltage of 30\,V. 
    The applied bias current as well as the  acquired dV/dI and DC voltage is shown as a function of time.
    (c)-(d) Decomposed interference pattern for both sweep directions recorded with the fast acquisition method. 
    The white line indicates the position where the line cut shown in panel (b) was taken.}
    \label{fig:fast_slow}
\end{figure}

In order to perform magnetic interference measurements, the bias current and external magnetic field can be swept between points. 
Such a measurement for $V_{\mathrm{BG}}=0\,\mathrm{V}$ is shown in Fig.~\ref{fig:fast_slow}(a). 
Using an acquisition method by applying the bias current as triangular-shaped pulses, the measurement time can be reduced. 
The results using this method are shown in Fig.~\ref{fig:fast_slow}(b). 
Here, the differential voltage and DC voltage are recorded simultaneously with the bias current pulse. 
This allows the magnetic field induced interference measurements to be acquired for both positive bias (see Fig.~\ref{fig:fast_slow}(c)) and negative bias (see Fig.~\ref{fig:fast_slow}(d)) in a shorter time.

To extract the switching current, data processing was conducted using a Savitzky-Golay filter, employing a window length of 61 and a polynomial order of 3, applied to data arrays of length 1024. 
With this filtered data, a threshold resistance ($R_{\mathrm{thresh}}$) was defined as the mean of 50 resistance values measured at high bias currents.\\
For current sweeps ranging from $-I_{\mathrm{max}}$ to $+I_{\mathrm{max}}$, the resistance $R(I)$ was evaluated sequentially from 0 to $+I_{\mathrm{max}}$ (-$I_{\mathrm{max}}$) to identify the switching currents $I_\mathrm{s}$ ($I_\mathrm{r})$ as the current points where the resistance exceeds $R_{\mathrm{thresh}}$.
If $R_{\mathrm{thresh}}$ was smaller than the minimum resistance value in the dataset, $I_\mathrm{s}$ and $I_\mathrm{r}$ were assigned a value of zero.\\
Further methodological details are provided in the scripts available at~\cite{zenodo}.

\clearpage

\section{Discussion of interference pattern}
Applying an out-of-plane magnetic field induces a phase difference between the two superconductors which leads to a modulation of the switching current $I_\mathrm{s}$ \cite{rowell1963sep}, as illustrated in Fig.~\ref{fig:fast_slow}(a).
Assuming a homogeneous current density, for a rectangular geometry as is used here, the maximum switching current $I_\mathrm{s}^\mathrm{m}$ follows a Fraunhofer type interference pattern given by
\begin{equation}
I_\mathrm{s}^m(\Phi)=I_\mathrm{s}\left|\frac{\sin\left(\frac{\pi\Phi}{\Phi_\mathrm{0}}\right)}{\frac{\pi\Phi}{\Phi_\mathrm{0}}}\right|
    \label{eq:fraunhofer}
\end{equation}
and depends on the total magnetic flux $\Phi=B_zA$ with the junction area $A=l\cdot w$, the magnetic flux quantum $\Phi_0=h/2e$ with the Planck constant $h$ and the elementary charge $e$, as well as the switching current $I_\mathrm{s}$ \cite{josephson1962jul}.
As the external magnetic field penetrates the superconductor for a certain distance $\lambda$ and thus alters the effective length of the junction, the period of the pattern can be used to determine this effective length $l_{\mathrm{eff}}$ \cite{london1934mar}.
With a period of 1.51\,mT, the effective length is determined to be $l_{\mathrm{eff}}\approx 603\,\mathrm{nm}$. Using the junction length of 250\,nm, as found in atomic force microscopy of the sample, the penetration depth is $\lambda\approx175\,\mathrm{nm}$ for each superconducting contact, comparable to literature values for niobium \cite{gubin2005aug}.

\begin{figure}[h]
    \centering
    \includegraphics[width=0.99\linewidth]{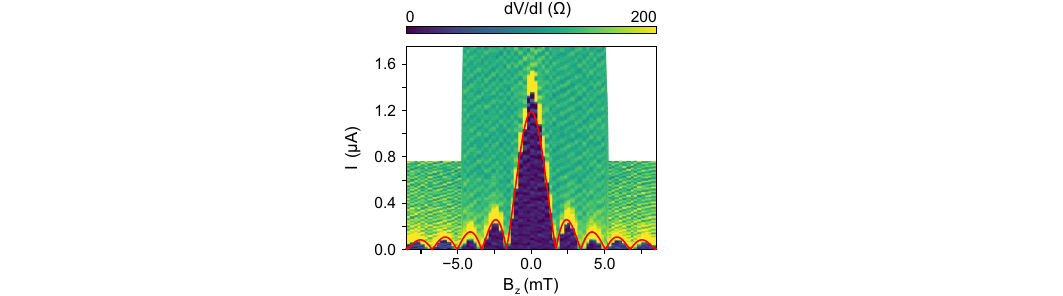}
    \caption{Measured interference pattern for $V_\mathrm{{BG}}=30\,\mathrm{V}$. 
    The red outline corresponds to a fit of the switching current to Eq.~(\ref{eq:fraunhofer}) with $\Phi=B_\mathrm{z}\cdot w\cdot l_{\mathrm{eff}}$.}
    \label{fig:fraunhofer_fit}
\end{figure}

\clearpage

\section{RCSJ model and determination of the capacitance}
The "resistively and capacitively shunted junction" (RCSJ) model as a fundamental theoretical framework is used to describe the dynamics of the Josephson junction \cite{stewart1968apr, mccumber1968jun}.
In this model, the junction is treated as an ideal Josephson element shunted by a resistor ($R$) and a capacitor ($C$), which account for dissipation and the junction's intrinsic capacitance, respectively. 
The model leads to a differential equation for the phase difference, $\varphi(t)$, across the junction, which is given by:
\begin{equation}
    I = I_\mathrm{s} \sin\varphi(t) + \frac{\hbar C}{2e} \frac{d^2\varphi(t)}{dt^2} + \frac{\hbar}{2eR} \frac{d\varphi(t)}{dt} + I_N,
\end{equation}
where $I$ is the total current through the junction, $I_\mathrm{s}$ is the switching current, $\hbar$ is the reduced Planck's constant, and $e$ is the elementary charge.
The first term represents the supercurrent, the second term accounts for the capacitive effects, and the third term represents the resistive current while the noise current $I_N$ accounts for finite temperature.
This equation describes the temporal evolution of the phase difference (and thus the voltage drop $V = \hbar/(2e) \, d\varphi/dt$) and provides insights into the junction's behavior under various conditions.

\begin{figure}[h]
    \centering
    \includegraphics[width=0.8\linewidth]{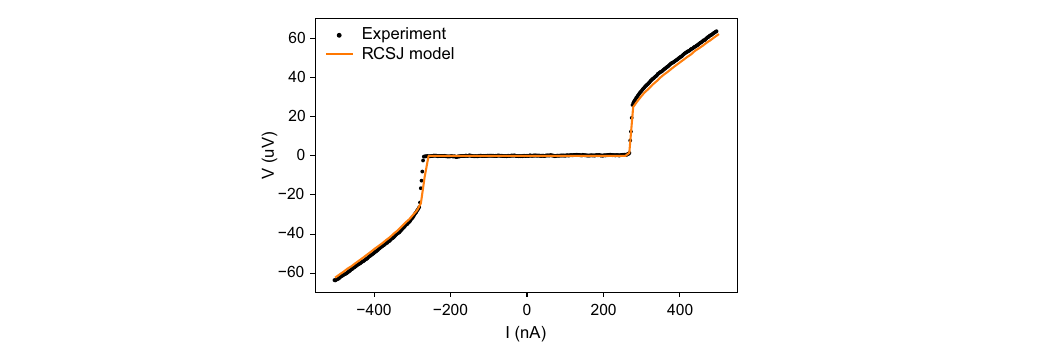}
    \caption{Measured and simulated V-I curve with junction parameters $R_\mathrm{s}=127\,\mathrm{\Omega}$, $C=0.1\,\mathrm{pF}$ and $I_\mathrm{s}=280\,\mathrm{nA}$.}
    \label{fig:rcsj_simulation}
\end{figure}

While the switching current and shunt resistance are directly accessible by the charge transport measurements, the junction's shunting capacitance remains to be determined by performing simulations of the RCSJ model.
We numerically solve the differential equation to simulate the V-I curves for a fixed, experimentally established switching current $I_\mathrm{s}$ and switching resistance $R_\mathrm{s}$ but varying capacitance $C$ and compare it with the experimental data, see Fig.~\ref{fig:rcsj_simulation}. 
We find that a capacitance of $C\approx0.1\,\mathrm{pF}$ leads to a good agreement between simulation and experiment and we use this value to determine the quality factors from the measured switching currents and resistances.

\clearpage

\section{$I_\mathrm{s}R_\mathrm{s}$ product}
\begin{figure}[h]
    \centering
    \includegraphics[width=0.99\linewidth]{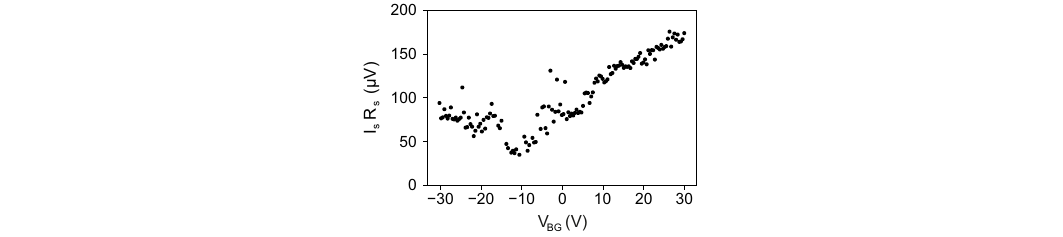}
    \caption{$I_\mathrm{s}R_\mathrm{s}$ product as function of back gate voltage.}
    \label{fig:IsRs}
\end{figure}

The $I_\mathrm{s}R_\mathrm{s}$ product provides insight into the junction's superconducting behavior, particularly in terms of its energy scale and coupling strength.
Depending on the applied gate voltage we get values ranging from $I_\mathrm{s}R_\mathrm{s}\approx 40\,\mathrm{\mu V}$ to $I_\mathrm{s}R_\mathrm{s}\approx 180\,\mathrm{\mu V}$, see Fig.~\ref{fig:IsRs}.

\section{Damping and quality factor}
\begin{figure}[h]
    \centering
    \includegraphics[width=0.99\linewidth]{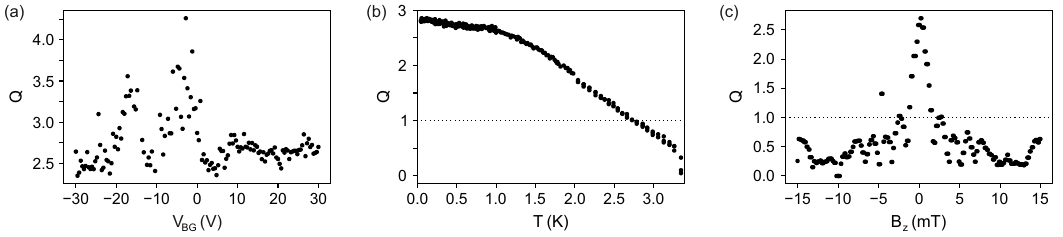}
    \caption{Quality factor $Q$ as function of external parameters gate voltage (a), temperature (b) and magnetic field (c).
    The junction's Q-factor, slightly modulated by gate voltage and consistently above one, transitions between weakly damped ($Q>1$) and highly damped ($Q<1$) regimes, depending on temperature and magnetic field.
    }
    \label{fig:Q}
\end{figure}

Within the RCSJ model the quality factor is defined as $Q=\sqrt{2eI_\mathrm{s}R_\mathrm{s}^2C/\hbar}$ and depends on the switching current $I_\mathrm{s}$, shunt resistance $R_\mathrm{s}$ and the junction capacitance $C=0.1\,\mathrm{pF}$. 
$Q\ll$1 describes a highly damped system and $Q\gg1$ describes a weakly damped system \cite{stewart1968apr, mccumber1968jun}.
To visualize the influence of external parameters, we extract the quality factor as function of gate voltage, temperature and magnetic field, see Fig.~\ref{fig:Q}.
While the gate voltage-dependent Q-factor shows a slight modulation, especially near the charge neutrality point, but is consistently greater than one, the Q-factor shows a transition between $Q>1$ and $Q<1$ depending on the temperature and magnetic field. 
Thus, the junction transitions between a weakly damped regime with $Q>1$ and a highly damped regime with $Q<1$, tunable by temperature and magnetic field.

\clearpage

\section{Additional Shapiro step measurements}
\subsection{Frequency dependence of the Shapiro steps}
\begin{figure}[h]
    \centering
    \includegraphics[width=0.99\linewidth]{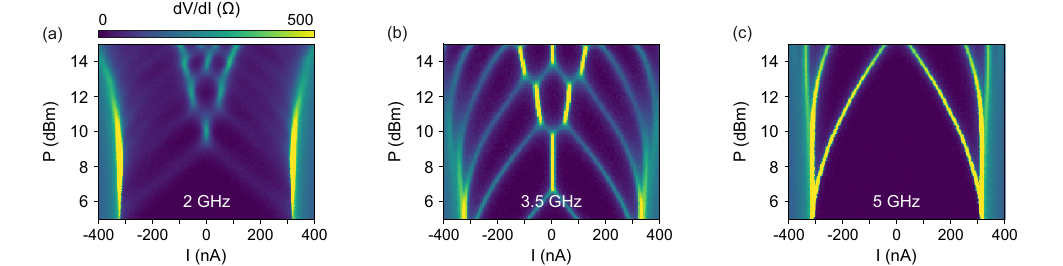}
    \caption{Shapiro maps at $V_{\mathrm{BG}}=30\,\mathrm{V}$ for 2\,GHz (a), 3.5\,GHz (b) and 5\,GHz (c). At 2\,GHz, the Shapiro steps are smeared out, whereas for 3.5\,GHz and 5\,GHz they become sharp. 
    However, for 5\,GHz, the first plateau does not become visible at powers up to 15\,dBm.}
    \label{fig:Shapiro_freq}
\end{figure}

The influence of the RF frequency on the Shapiro steps was examined to determine the frequency at which the radiation is most efficient. 
In addition to the measurement taken at 4\,GHz (see Fig.~2 of the main manuscript), Shapiro steps were examined at 2\,GHz, 3.5\,GHz, and 5\,GHz (see Figs.~\ref{fig:Shapiro_freq}(a)-(c)). 
For lower frequencies the steps become smeared out but the overall shape of the pattern remains the same.
At 5\,GHz, the plateaus become sharper compared to the lower frequencies (see Fig.~\ref{fig:Shapiro_freq}(c)); however, even the first Shapiro step does not develop within the available power range. Therefore, 4\,GHz is determined to be the most suitable drive frequency.

\subsection{Shapiro steps at different damping regimes}
\begin{figure}[h]
    \centering
    \includegraphics[width=0.99\linewidth]{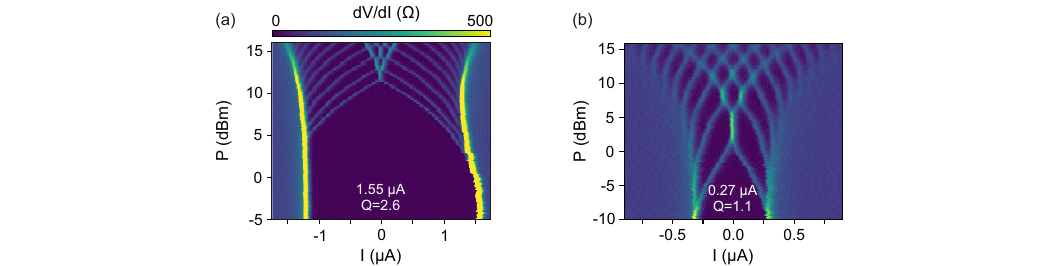}
    \caption{(a) and (b) Shapiro step pattern measured at $V_{\mathrm{BG}}=30$\,V and different out-of-plane magnetic fields. 
    At $-0.15$\,mT in panel (a), the switching current of 1.55\,µA is larger than the switching current at 0.85\,mT in panel (b), which corresponds to 0.27\,µA. 
    This results in two different damping regimes, with (a) being underdamped and (b) being intermediately damped.}
    \label{fig:Shapiro_damp}
\end{figure}

To investigate the influence of the junction's damping regime on the pattern of the Shapiro steps, the measurements are performed for different values of the quality factor $Q$.
The value of $Q$ is controlled through application of a perpendicular magnetic field. 
The damping influences the appearance of the Shapiro steps as depicted in Figs.~\ref{fig:Shapiro_damp}(a) and (b) where the highly damped case with $Q=2.6$ and the intermediately damped case with $Q=1.1$ is shown, respectively.
For $Q=2.6$ and $I_{\mathrm{s}}=1.55$\,µA, the Shapiro steps are limited to bias current values up to $I_{\mathrm{s}}$. 
In contrast, for $Q=1.1$ and $I_{\mathrm{s}}=0.27$\,µA, the Shapiro steps extend well beyond $I_{\mathrm{s}}$. 
These characteristics in the moderately damped regime qualitatively resemble Shapiro maps observed in graphene Josephson junctions documented in literature \cite{larson2020oct,kalantre2020apr}.

\subsection{Half integer switching}
\begin{figure}
    \centering
    \includegraphics[width=0.99\linewidth]{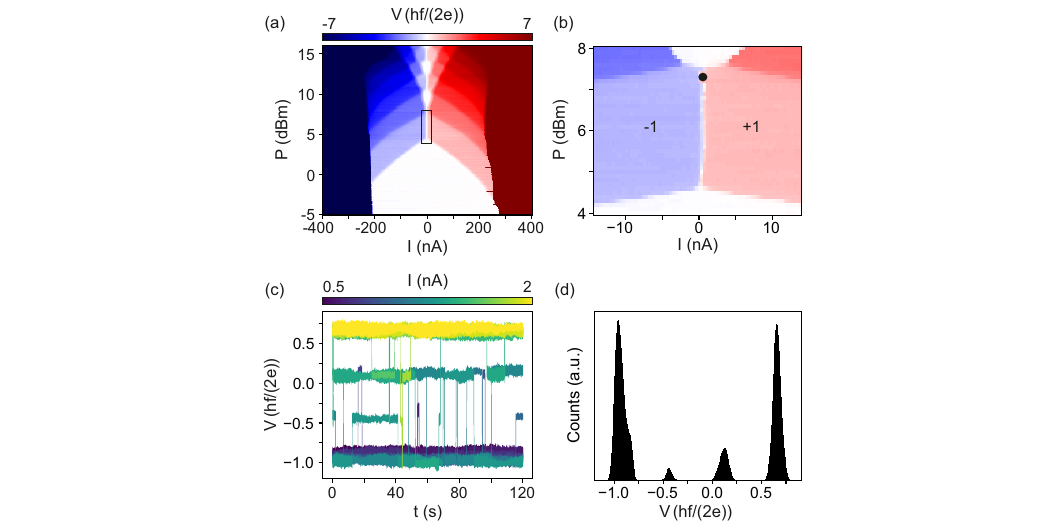}
    \caption{
        (a) Shapiro step measurement taken at $V_{\mathrm{BG}}=0\,\mathrm{V}$ and $f=4\,\mathrm{GHz}$.
        (b) Zoom-in on the $\pm1$ transition. 
        The black dot indicates the examined position for (c) and (d).
        (c) Time dependent DC voltage for different bias currents where switching between four distinct states is observable. 
        The occupation of the half integer steps lasts up 20\,s.
        (d) Histogram of DC voltages in units of $hf/2e$ showing the half integer steps.
    }
    \label{fig:Shapiro_halfinteger}
\end{figure}

Two neighboring Shapiro plateaus represent a two-level system, and thus, stochastic switching between the two levels might be observable when the bias current and the output RF power are tuned exactly at the crossover between two plateaus. 
To observe this, the Shapiro step measurement taken at 0\,V back gate voltage in Fig.~\ref{fig:Shapiro_halfinteger}(a) was examined and a zoom-in on the border between the +1 and -1 plateaus was investigated (see Fig.~\ref{fig:Shapiro_halfinteger}(b)).
The black dot in panel (b) indicates the RF power in Figs.~\ref{fig:Shapiro_halfinteger}(c) and (d).
Notably, in Fig.~\ref{fig:Shapiro_halfinteger}(d), it is evident that switching occurs not only between the +1 and -1 plateaus but also at fractional values in between.
The measured DC voltage is shown as a function of time and applied bias current. 
The distance between the states is approximately 0.5 $hf$/2$e$. 
These fractional states appearing between the +1 and -1 plateaus exhibit lifetimes as long as 20 seconds.

\clearpage

\section{Evolution of interference pattern subject to in-plane magnetic fields}
\begin{figure}[h]
    \centering
    \includegraphics[width=0.99\linewidth]{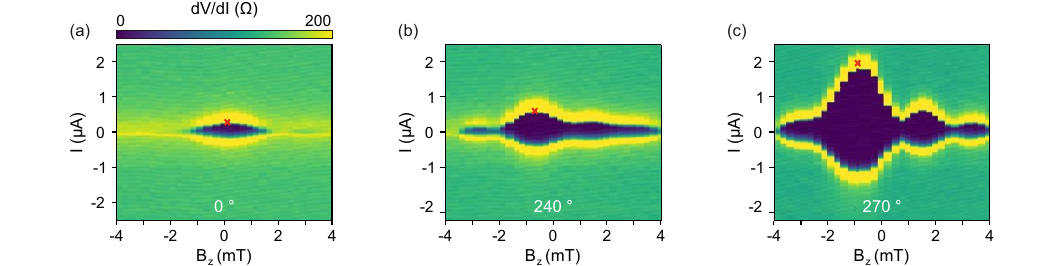}
    \caption{Interference measurements acquired at $B_{||}$=200\,mT for $\varphi$=0\,°, 240\,° and 270\,° are shown in (a), (b) and (c), respectively. 
    The red cross indicates the extracted maximum of the switching current which was used in the polar plot in the main manuscript to correct for the tilt of the sample.}
    \label{fig:Bip_correction}
\end{figure}

As the sample is slightly tilted inside the sample holder, in-plane fields result also in non-negligible out-of-plane fields which need to be corrected. 
This is evident in the acquired Fraunhofer-like interference measurements, where the zeroth maximum is shifted with respect to $B_\mathrm{z}=0\,\mathrm{mT}$ in the presence of an applied in-plane magnetic field of $B_{||}=200\,\mathrm{mT}$ (see Figs.~\ref{fig:Bip_correction}(a),(b) and (c)). 
To correct for this effect, only the maximum switching current (red crosses in Figs.~\ref{fig:Bip_correction}(a),(b) and (c)) is taken into account for the polar plot shown in the main manuscript.

\begin{figure}[h]
    \centering
    \includegraphics[width=0.99\linewidth]{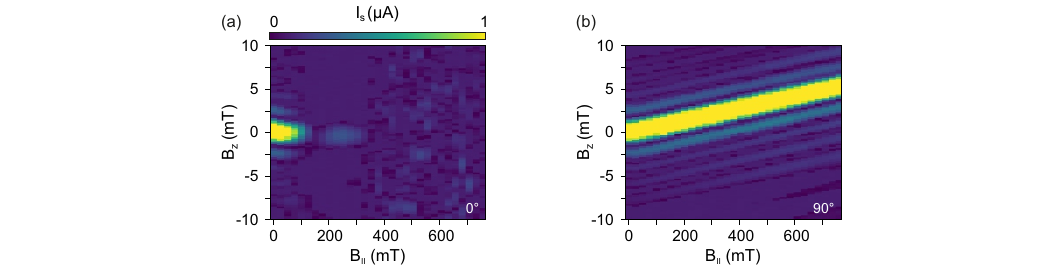}
    \caption{The extracted maximum switching current $I_\mathrm{s}$ is shown as a function of applied perpendicular magnetic field $B_z$ and in plane magnetic field $B_{||}$ with an angle of 0° (90°) with respect to the direction of current direction in a (b). 
    While the interference pattern is strongly suppressed with increasing in plane field in (a), its zeroth maximum is only shifted towards higher $B_z$ in panel (b). 
    Here, the data is not corrected for $B_z$-offsets.}
    \label{fig:Bx_Bz_Ic}
\end{figure}

In addition to interference measurements acquired at fixed amplitudes of $B_{||}$ and various in-plane angles, the amplitude of $B_{||}$ was varied at fixed angles of the in-plane magnetic field, see Fig.~\ref{fig:Bx_Bz_Ic}. 
The supercurrent pattern is increasingly suppressed at $\varphi=0$\,° with increasing $B_{||}$ whereas for $\varphi=90$\,° the interference pattern shifts linearly but the supercurrent is not suppressed. 
The linear shift, which we ascribe to a slight misalignment of the sample plane with the vector magnet axes, is used to extract a correction angle of $0.41\pm0.01$\,° with respect to the magnetic field axes, and we correct for this angle when sweeping the in-plane magnetic field to obtain the data shown in Fig.~3 of the main text.

We note that we extract from our AFM image an in-plane angle of 60\,° of the junction with respect to the magnet x-axis. 
We estimate an accuracy of $\pm5$\,° based on our sample mounting accuracy. 
This angle has been corrected throughout our study for simplicity, so that the x-axis now lies along the sample axis, as shown in Fig.~1 of the main text.

\clearpage

\section{Determination of the in-plane magnetic field induced switching current minima}

\begin{figure}[h]
    \centering
    \includegraphics[width=0.99\linewidth]{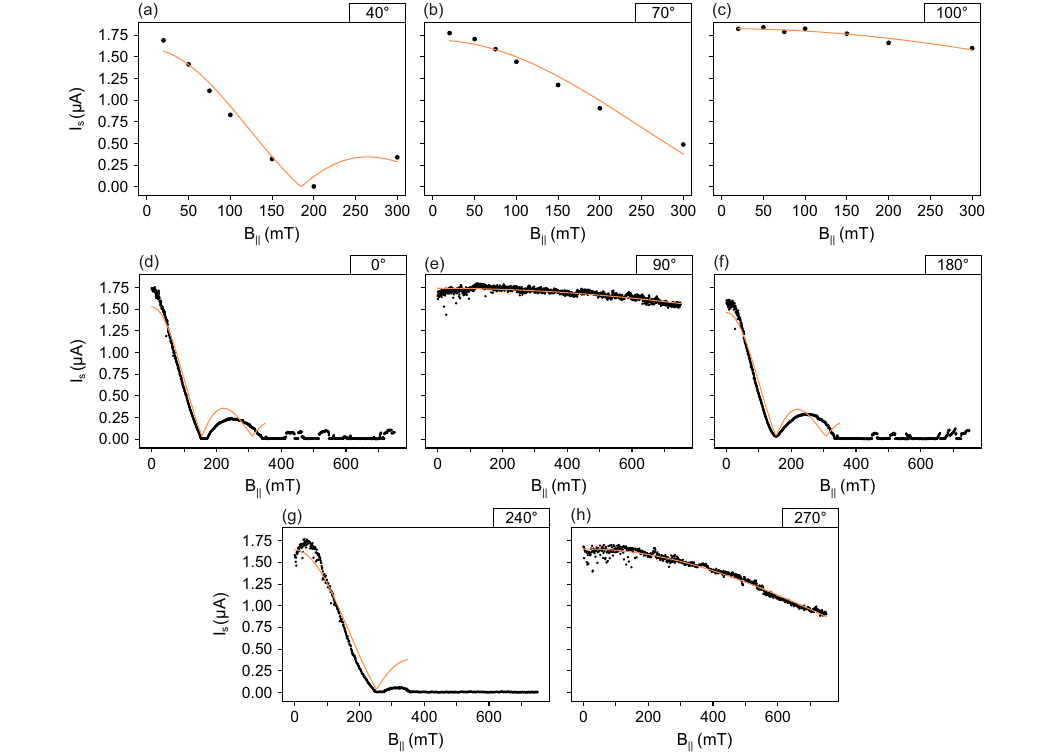}
    \caption{Switching current as function of in-plane magnetic field for different angles $\varphi$. 
    The magnetic field of the first minimum is determined by fitting a sinc function (orange lines).
    (a) - (c) and (d) - (h) correspond to the datasets presented in Fig.~4(b) and Fig.~4(e) of the main text, respectively.}
    \label{fig:in_plane_decay}
\end{figure}

To determine the in-plane magnetic field at which the switching current hosts a minimum, we plot the extracted switching current as a function of the in-plane magnetic field for various angles (same dataset as in Figs.~4(b) and 4(e) of the main text) and fit the model $I_\mathrm{s}=|a\,\mathrm{sinc}(b\, B_{||})|$ to the data, see Fig.~\ref{fig:in_plane_decay}. 
Thus, the magnetic fields corresponding to the first minimum are given by $B_{||}^1 = \pi/b$.

After all, the assumption of a simple rectangular geometry with a homogeneous supercurrent density is unlikely to be true in a mesoscopic device like ours. 
In fact, the apparent ease of fitting the data with this simple model is somewhat surprising. 
We speculate that the large back gate voltage we chose, with correspondingly large electron density, leads to a somewhat homogeneous supercurrent density leaving the model somewhat applicable. 
However, a detailed fitting or explanation based on this model, beyond extracting and interpreting a simple coarse feature like the first supercurrent minimum, seems pointless, as any added complexity such as disorder or indeed spin-related effects will cause deviations that are not captured by our model. 
It is for this reason that fitted supercurrent minima relaying on significant extrapolation should be treated with caution. 
In particular, the outlying point at 90 degrees is not to be taken seriously in our opinion, as only an initial 15\% decay can be used for the fitting procedure (see Fig. \ref{fig:in_plane_decay}), and this is why we have not included it. 

\begin{figure}[h]
    \centering
    \includegraphics[width=0.99\linewidth]{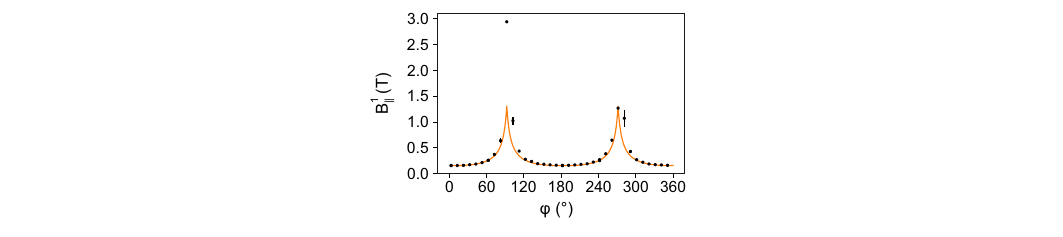}
    \caption{Full dataset of Fig.~4f. 
    In-plane magnetic field of the first supercurrent minimum $B^1_{||}$ as function of in-plane angle. 
    Solid line represents a fit to the geometrical model described in the main text.
    We consider the point at 90 degrees an outlier and do not include it in the fit. }
    \label{fig:in_plane_full}
\end{figure}